\documentclass[aps,pre,superscriptaddress,twocolumn,amsmath,amssymb,showpacs]{revtex4}
\usepackage{amsmath}
\usepackage{graphicx}
\usepackage{amssymb}

\usepackage{dcolumn}
\usepackage{bm}

\makeatletter

\begin{document}

\title{Entropic transport: Kinetics, scaling and control mechanisms}

\author{D. Reguera}
\affiliation{Department de F\'isica Fonamental,
  Facultat de F\'isica,
  Universidad de Barcelona,
  Diagonal 647, E-08028 Barcelona, Spain}

\author{G. Schmid}
\affiliation{Institut f\"ur Physik,
  Universit\"at Augsburg,
  Universit\"atsstr. 1,
  D-86135 Augsburg, Germany}

\author{P. S. Burada}
\affiliation{Institut f\"ur Physik,
  Universit\"at Augsburg,
  Universit\"atsstr. 1,
  D-86135 Augsburg, Germany}

\author{J. M. Rub\'i}
\affiliation{Department de F\'isica Fonamental,
  Facultat de F\'isica,
  Universidad de Barcelona,
  Diagonal 647, E-08028 Barcelona, Spain}

\author{P. H\"anggi}
\affiliation{Institut f\"ur Physik,
  Universit\"at Augsburg,
  Universit\"atsstr. 1,
  D-86135 Augsburg, Germany}

\date{\today}

\begin{abstract}
\noindent We show that transport in the presence of {\it entropic}
barriers exhibits peculiar characteristics which makes it distinctly
different from that occurring through energy barriers. The
constrained dynamics yields  a scaling regime for the particle
current and the diffusion coefficient in terms of the ratio between
the work done to the particles and available thermal energy. This
interesting property, genuine to the entropic nature of the
barriers, can be utilized to effectively control  transport through
quasi one-dimensional structures in which irregularities or
tortuosity of the boundaries cause entropic effects. The accuracy of
the kinetic description
has been corroborated by simulations. Applications to different
dynamic situations involving entropic barriers are
outlined.
\end{abstract}

\pacs{02.50.Ey, 05.40.Jc, 05.60.Cd}

\maketitle

Transport through quasi-onedimensional structures as pores, ion
channels and zeolites is ubiquitous in biological and
physico-chemical systems and constitute a basic mechanism in
processes as catalysis, osmosis and particle separation
\cite{hille, zeolites, Chou, liu, siwy, berzhkovski}.  A common
characteristic of these systems is the confinement arising from the
presence of boundaries which very often exhibit an irregular
geometry. Variations of the shape of the
structure along the propagation direction implies changes in the number
of accessible states of the particles. Consequently, entropy is spatially
varying, and the system evolves through entropic barriers, which controls
the transport, promoting or hampering the transfer of mass and energy
to certain regions.
Motion in the system can be induced by the presence of external
driving forces supplying the particles with the energy necessary to
proceed. The study of the
kinetics of the entropic transport, the properties of transport
coefficients in far from equilibrium situations and the possibility
for transport control mechanisms are objectives of major
importance in the dynamical characterization of those systems.

Our purpose in this Letter is to demonstrate that entropic transport
exhibits striking features, sometimes counterintuitive, which are different
from those observed in the more familiar case with energy barriers \cite{hanggi}.
We propose a general scenario describing the dynamics through
entropic barriers and show the existence of a scaling regime for the
current of particles and the effective diffusion coefficient. The
presence of this regime might have important implications in the
control of transport.

\begin{figure}[t]
\includegraphics{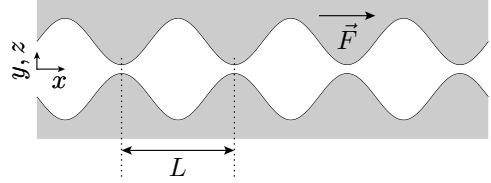}
\caption{Schematic diagram of the tube confining the
  motion of the biased Brownian particles. The half width $\omega$ is a
  periodic function of $x$ with periodicity $L$.}\label{fig:sketch}
\end{figure}

{\it Entropic transport.} - \hspace*{0.05cm} The origin of the
entropic barriers can be inherent to the intimate nature of the system
or may emerge as a consequence of a coarsening of the description
employed. A typical example presents the motion of a Brownian
particle in an enclosure of varying cross-section. This basic
situation constitutes the starting point in the study of transport
processes in the type of confined systems that are very often
encountered at sub-cellular level, nanoporous materials and in
microfluidic applications. As shown in Ref. \cite{Zwanzig}, the
complicated boundary conditions of the diffusion equation in
irregular channels can be greatly simplified by introducing an
entropic potential
that accounts for the reduced space accessible
for the diffusion of the Brownian particle. The resulting kinetic
equation describing the evolution of the probability distribution,
is known as the Fick-Jacobs equation \cite{Zwanzig, Jacobs} and
constitutes an approximation to the full dynamics. The validity of
that equation has only been analyzed for diffusion in the
absence of a driving force in whose case many of the transport
processes previously mentioned could not take place. This is so,
since thermal diffusion alone may not be able to induce transitions
of the particles through the entropic barrier.

In typical transport processes through pores or channels, motion of
the suspended particles is induced by application of an external
driving force $F$ that is directed along their axis. The over-damped
dynamics of a biased Brownian particle within the tube (see
Fig.~\ref{fig:sketch}) then reads:
\begin{align}
  \label{eq:langevin}
  \eta\, \frac{\mathrm{d}\vec{r}}{\mathrm{d} t}=\vec{F} + \sqrt{\eta\, k_\mathrm{B}\,T}\, \vec{\xi}(t)\, ,
\end{align}
where $\eta$ is the friction coefficient of the particle,
$k_\mathrm{B}$ the Boltzmann constant, $T$ the temperature,
$F=|\vec{F}|$ a constant force in $x$-direction and $\vec{\xi}(t)$
is Gaussian white noise with zero mean and correlation function:
$\langle \xi_{i}(t)\,\xi_{j}(t') \rangle = 2\, \delta_{ij}\, \delta(
t - t')$ for $i,j = x,y,z$. The reflecting boundary conditions
ensure the confinement of the dynamics within the tube.

{\it Reduction of the dimensionality.} - \hspace*{0.05cm} As
mentioned previously, the dynamics of the particles along the axis
of the 3D tube or a 2D channel (see Fig.~\ref{fig:sketch}) can be
recast into the {\it Fick-Jacobs equation}; i.e.,

\begin{align}
\label{eq:fickjacobs}
\frac{\partial P}{\partial t}=D_{0}\frac{\partial}{\partial
  x}\left(s(x)\frac{\partial}{\partial
    x} \frac{P}{s(x)}\right)
\end{align}
obtained from the 3D (or 2D) Smoluchowski equation after elimination
of $y$ and $z$ coordinates by assuming equilibrium in the orthogonal
directions. Here $P(x,t)$ is the probability distribution function,
$D_0$ the diffusion coefficient, and $s(x)$ is the cross-sectional area
for a (3D) tube or the width for a (2D) channel.
This description is in principle valid for
$\left|\omega'(x)\right|\ll1$, where $\omega(x)$ is the radius of the tube (or the
halfwidth of the channel in 2D) and the prime refers to the first
derivative. It has been shown that the
introduction of a $x$-dependent diffusion coefficient considerably
improves the accuracy of the kinetic equation extending its validity
to more winding structures \cite{Zwanzig,Reguera}. The expression
\begin{align}
  \label{eq:diffusionconst}
  D(x)=\frac{D_{0}}{(1+\omega'(x)^{2})^{\alpha}}\, ,
\end{align}
where $D_{0} = k_\mathrm{B}\,T / \eta$ and $\alpha=1/3,1/2$ for two and three
dimensions, respectively, was shown to appropriately account for the
curvature effects \cite{Reguera}.

In the presence of a constant force $F$ along the direction of the
tube the Fick-Jacobs equation can be recast into the following expression \cite{Reguera}

\begin{align}
\label{eq:fickjacobs_ours}
\frac{\partial P}{\partial t}=\frac{\partial}{\partial
  x}\left(D(x)\frac{\partial P}{\partial
    x}+\frac{D(x)}{k_\mathrm{B}\,T}\frac{\partial A(x)}{\partial x}P\right)
\end{align}
which defines a free energy
  $A(x) := E - T\, S = -F\, x - T\, k_\mathrm{B}\, \ln  h(x)$,
where $E=-Fx$ is the energy, $S=k_\mathrm{B} \,\ln h(x)$  the
entropy, $h(x)$ the dimensionless width $2\,\omega(x)/ L$
in 2D, and the dimensionless transverse cross-section
$\pi\, [\omega(x)/L]^{2} $ of the tube in 3D.
For a symmetric channel with periodicity $L$, the free energy assumes the form of a periodic tilted potential.

{\it Universal scaling for the particle current and effective
  diffusion.}
- \hspace*{0.05cm}
The
key quantities in transport through quasi one-dimensional structures
are the average particle current, or equivalently the nonlinear
mobility, and the effective diffusion coefficient. While in the case of an energy barrier,
the driving force $F$ and
the temperature $T$ are two independent variables, for
entropic transport, both current and effective diffusion are controlled by
a universal scaling parameter:
\begin{align}
  \label{eq:scaling}
  f:= \frac{F\, L}{k_\mathrm{B}\,T}\, .
\end{align}
For the average particle current and the nonlinear mobility $\mu(f)$ we
find an expression similar to the Stratonovich
formula \cite{Reimann}

\begin{subequations}
  \label{eq:nonlinearmobility}
\begin{align}
  \mu(f) :=  \frac{\langle \dot{x} \rangle}{F} = \frac{1}{\eta}\, \frac{1-\exp(-f)}{\int_{0}^{1}\,
      \mathrm{d}z \, I(z, f)}\, f^{-1} \, ,
\end{align}
where
\begin{multline}
  \label{eq:integrals}
  I(z, f) := \left[1 + \left( \hat{\omega}'(
        z) \right)^{2}\right]^{\alpha}\, \hat{h}^{-1}( z) \\ \times  \exp(- f\, z)\,  \int_{z -
    1}^{z} \mathrm{d}\tilde{z} \, \hat{h}(\tilde{z})\, \exp(f\, \tilde{z}) \, ,
\end{multline}
\end{subequations}
depends only on the dimensionless variable $z=x/L$, the scaling parameter $f$ and the shape of the tube given
in terms of the dimensionless half width $\hat{\omega}(z):=w(x)/L$
and its first derivative. Here $\hat{h}(z):=h(x)$.

The effective diffusion coefficient could be expressed in terms of moments of the
first passage time for a Brownian particle arriving at $x_{0}+L$
while starting out from $x_{0}$ \cite{Reimann}.
A detailed analysis shows that the effective diffusion
coefficient also
scales with $F\, L/k_\mathrm{B}\,T$ as:
\begin{subequations}
  \label{eq:effectivediffusion}
  \begin{align}
    \frac{D_\mathrm{eff}}{D_{0}}= \frac{\int_{0}^{1}\mathrm{d}z\,\int_{z-1}^{z}\mathrm{d}\tilde{z}\,
      {\cal N}(z, \tilde{z}, f)}{\left[\int_{0}^{1}\mathrm{d}z\,
        I(z,f)\right]^{3}}\, ,
  \end{align}
  with
  \begin{multline}
    {\cal N}( z, \tilde{z}, f):=
    \left(\frac{1+\left(\hat{\omega'}(z)\right)^{2}}{1+\left(\hat{\omega}'
    (\tilde{z})\right)^{2}}\right)^{\alpha}\,
    \frac{\hat{h}(\tilde{z})}{\hat{h}(z)}\\
    \times \left[I(\tilde{z},f)\right]^{2} \, \exp(-f\, z + f\,  \tilde{z})\, .
  \end{multline}
\end{subequations}

\begin{figure}[t]
\includegraphics{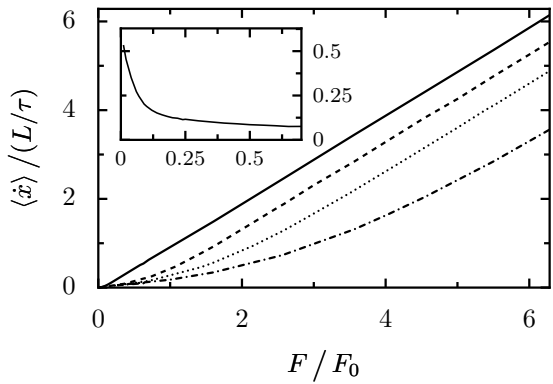}
\caption{Numerically determined force dependence of particle current
for a symmetric two-dimensional tube with the shape defined by the
half width $\omega(x)= [\sin(2\pi x/L)+ 1.02]/(2\pi)$, $L=1$ and for the
values of $T\big/T_{\text{room}}$: $0.01$ (solid line), $0.1$ (dashed
line), $0.2$ (dotted line) and $0.4$ (dash-dotted line). The inset
depicts the dependence of the particle current $\langle \dot{x}
\rangle / (L/\tau)$ on the dimensionless temperature $T/T_\text{room}$
for the force value: $F/ F_{0} = 0.628$. Contrary to the case of energetic barriers,
the particle current declines with increasing
temperature.}\label{fig:currentvsforce}
\end{figure}

\begin{figure}[ht]
\includegraphics{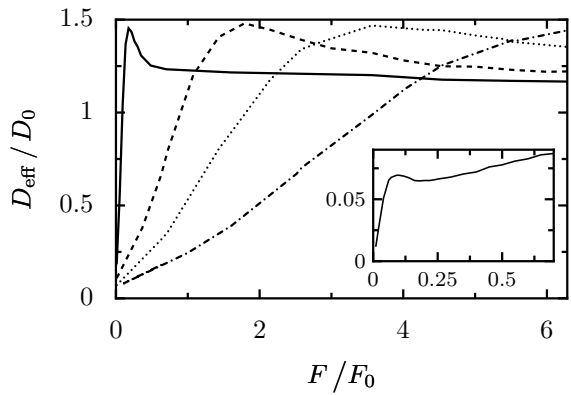}
\caption{The
  effective diffusion coefficient versus the external bias. The
  parameters for the various lines correspond to those detailed in
  Fig.~\ref{fig:currentvsforce}.  The inset depicts
  the effective diffusion coefficient $D_{\text{eff}} / (L^{2}/\tau)$
  vs. dimensionless temperature
  $T/T_{\text{room}}$.}\label{fig:diffusionvsforce}
\end{figure}

{\it Numerical simulations.} - \hspace*{0.05cm}
A model of a 2D periodic channel is sketched in
Fig.~\ref{fig:sketch},
the shape is described by
$\omega(x) = a\sin(2\pi x/L)+b$. Here, $a$ is the parameter that controls the
slope of the walls, the width of the channel is $2\, \omega(x)$, and
the width at the bottleneck is $2\,(b-a)$.

The scaling behaviors, predicted above, have been corroborated by Brownian
dynamic simulations performed by  integration
of the dimensionless Langevin equation, which is equivalent to Eq.~\eqref{eq:langevin}, within the
stochastic Euler-algorithm. Therefore lengths are scaled by the
periodicity $L$ of the tube, time by $\tau:=L^{2}\eta
/(k_{\mathrm{B}}\, T_{\text{room}})$ -- the corresponding
characteristic diffusion time at room temperature
$T_{\text{room}}$ --   and the force by
$F_{0}:=\eta\, L/\tau$.
The mean velocity in $x$-direction, $\langle \dot{x}
\rangle=\lim_{t\to \infty} x(t)/t$, and the corresponding effective diffusion
coefficient, $D_{\mathrm{eff}} = 1/2\, \lim_{t \to \infty} (\langle
x^{2}(t) \rangle - \langle x(t) \rangle^{2} )/t$,  are obtained as
an average over $3\cdot 10^{4}$ trajectories.

\begin{figure}[floatfix]
\includegraphics{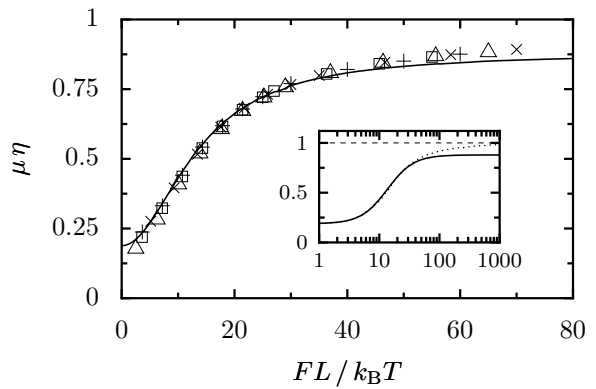}
\caption{
  Graph for the scaled nonlinear mobility. In the Langevin simulation the different symbols correspond to
  different values of $T/T_{\text{room}}$: $0.01$ (crosses), $0.1$ (pluses), $0.2$ (squares), $0.4$ (triangles).
  The relative error of the simulation results is $0.01$.
  The
  Fick-Jacobs results, Eq.~\eqref{eq:nonlinearmobility}, correspond to the solid lines.
  The inset depicts the long range behavior (the dotted
  line depicts the numerical
  results).
  The numerical values for the scaled nonlinear mobility
  approach, in the limit $f\to\infty$, to the value
  $1$ (dashed horizontal line). }\label{fig:scaledcurrent}
\end{figure}

\begin{figure}[floatfix]
\includegraphics{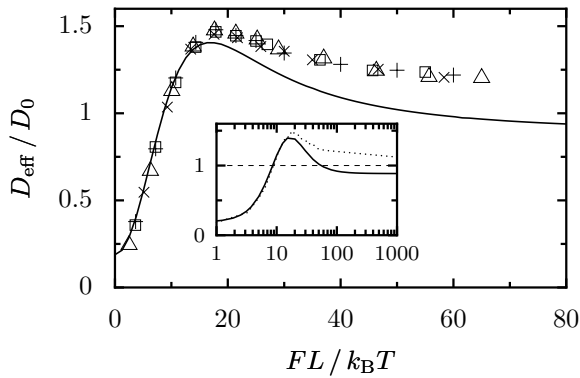}
\caption{Same as in Fig.~\ref{fig:scaledcurrent}, but for the
  effective diffusion
  coefficient.
  The relative error of the simulation results is $0.1$.
  The
  Fick-Jacobs results, Eq.~\eqref{eq:effectivediffusion}, correspond to the solid lines.
  The inset depicts the long range behavior (the dotted
  line depicts the numerical
  results). The numerical values for the scaled nonlinear mobility
  approach, in the limit $f\to\infty$, to the value
  $1$ (dashed horizontal line).}\label{fig:scaleddiffusion}
\end{figure}

Results for the particle current
and the effective diffusion coefficient as a function of the applied force for
the case $a=1 {\big/} (2\, \pi)$, $b=1.02 \big/ (2\, \pi)$ and $L=1$ are presented in
Fig.~\ref{fig:currentvsforce} and Fig.~\ref{fig:diffusionvsforce} for
different values of the noise strength (i.e. the temperature). The particle current  increases
monotonically with the  force, but \emph{decreases} upon
increasing the level of noise. The effective
diffusion coefficient exhibits a non-monotonic behavior with the
appearance of a peak which becomes more pronounced at low noise levels
(see Fig.~\ref{fig:diffusionvsforce}).
When both quantities are represented as a function of the scaling
parameter $f$ (see
Fig.~\ref{fig:scaledcurrent} and Fig.~\ref{fig:scaleddiffusion}) all
curves collapse to the scaled solution which evidences the excellent agreement
of
simulations results with the scaling behavior predicted for those
quantities. Therefore, whereas in the case of transport through
energetic barriers the force (or tilt) and the temperature are two independent
parameters, the entropic transport is controlled by a single parameter
$f$. Another important results shown in Fig.~\ref{fig:diffusionvsforce} is the presence of
a peak in the diffusion and the fact that the effective diffusion
can be much larger than bulk diffusion. Thus the phenomenon of enhancement
of the diffusion, linked to the dynamics of particles in periodic
tilted energetic potentials, also takes place when barriers hindering
the transport have an entropic nature.

In Fig.~\ref{fig:scaledcurrent} and Fig.~\ref{fig:scaleddiffusion} we
have also represented the nonlinear mobility and effective diffusion
coefficient predicted by Eq.~\eqref{eq:nonlinearmobility} and \eqref{eq:effectivediffusion}
obtained from the Fick-Jacobs
equation. At low values of the scaling variable $f$ the results match
perfectly with the simulations whereas deviations occur at higher values
of $f$. The scaled nonlinear
mobility and the effective diffusion coefficient approximates for
$f\to\infty$ values different from the value $1$.
The accuracy of the Fick-Jacobs description worsens at large $f$ because
the assumption of equilibration in the transverse direction, which
supports the elimination of the y,z coordinates, fails at high values
of the applied force. The agreement substantially improves when the
shape of the tube does not change too fast, i.e. when $\left|\omega'(x)\right|$
is smaller, which can be achieved for instance by increasing the period
$L$ of the shape oscillations of the channel. In situations where the
roughness of the channel is not very extreme, the Fick-Jacobs description
provides a very good approximation to the transport for values of
the external work of some tens of $k_{\mathrm{B}}\, T$'s. In fact, that is the
range of energies relevant to most transport processes in biological
systems.

The peculiar behavior of nonlinear mobility and effective diffusion
coefficient as a function
of temperature is depicted in the insets of Fig.~\ref{fig:currentvsforce}
and \ref{fig:diffusionvsforce}. Contrary to the case of an energetic
barrier, the particle current \emph{decreases} upon increasing the temperature.
In the presence of energetic barriers, the temperature facilitates
the activation (the overcoming of the barriers) and thus tends to
increase the particle current. However, when transport is controlled by entropic
factors, the temperature dictates the strength of the entropic potential,
and thus an increase of temperature leads to a reduction of the particle current.
The effective diffusion coefficient as a function of the temperature also manifest
a striking behavior with the presence of a peak, and the existence
of a range of temperatures where the effective diffusion coefficient decreases
upon increasing the temperature. It is important to remark that, since
the transport characteristics scale as $FL/k_\mathrm{B}\,T$, the peculiar
regimes can be obtained not only by changing the temperature but also
by modifying the strength of the force.

{\it Applications.} - \hspace*{0.05cm}
An example in which the entropic nature of the transport becomes more
evident is the case of micro and nanoporous materials, such as zeolites.
These materials have a regular structure with channels of different
width and well-defined geometry. This peculiar structure confers them
an outstanding ability to act as molecular sieves, that is currently
exploited in chemically clean separation of mixtures, ion exchange
and petrochemical cracking. Driven by their economic and scientific
importance, these materials have been studied extensively experimentally
and more recently by computer simulations.
For instance, the diffusion has been found to
decrease with temperature in some range of temperatures \cite{churring}; and
the existence of an optimal value of the diffusion as a function of
the temperature has also been observed \cite{dubbel}.
In fact,
the dependence of the effective
diffusion coefficient
on temperature reported in Ref. \cite{dubbel} behaves just as the one
predicted here with Fig.~\ref{fig:diffusionvsforce}.
Finally, values of diffusion coefficients higher than the bulk, consistent with the
phenomenon of diffusion enhancement predicted by our model,  have also been
reported \cite{yashonath}. Our simple model thus accounts for all these behaviors and
 shows that they are
not specific of a particular zeolite structure but they arise from
the entropic nature of the transport.

{\it Conclusions.} - \hspace*{0.05cm}
In summary, we have shown that transport phenomena in systems in the
presence of entropic barriers exhibit some features radically different
from conventional transport through energetic barriers. The effect
of confinement can be recast in terms of an entropic potential, and
the dynamics of the system can be accurately described by means of
the Fick-Jacobs equation. We have shown the existence of a scaling
regime in the dynamics. The particle current and the effective
diffusion coefficient are controlled by a single parameter $f$ that
measures the relative importance of the external work done to the particle and the
thermal energy. The scaling in $f$ thus opens up the possibility
of tuning and controlling the efficiency of transport in confined
systems by a proper combination of temperature and applied field. In situations
in which the temperature can
only be varied in a very limited range, as frequently occurs in biological
systems, the existence of scaling implies
that the same transport regime can be accomplished by the
application of an external force. The analysis presented could be applied to a
wide variety of situations, such as biological transport through
ion channels and membrane pores, or the portage in molecular sieves
or polymer gels, where entropic effects  play
a very important role.

We thank P. Reimann for fruitful discussions. This work has been supported by the
DGiCYT under Grant No BFM2002-01267, the DAAD,
ESF STOCHDYN project, the Alexander von Humboldt Foundation,
the DFG via
research center, SFB-486, project A10
 and via the project
no. 1517/26-1.

\end{document}